\documentclass[10pt,notitlepage,nofootinbib,preprintnumbers,showpacs]{revtex4-1}
\usepackage{color,graphicx,epsfig} 
\usepackage{ifpdf}
\usepackage{amsmath} 
\usepackage{amssymb}
\usepackage{bm}

\usepackage{color}
\definecolor{nicered}{rgb}{0.7,0.1,0.1}
\definecolor{nicegreen}{rgb}{0.1,0.5,0.1}

\usepackage[english]{babel}
\usepackage{slashed}
\usepackage{multirow}

\pdfoutput=1
\newcommand {\E}[1]{\times 10^{#1}}	
\newcommand {\e}[1]{\mathrm{~#1}}       
\newcommand{\mc}[1]{\mathcal{#1}}

\newcommand{\mrm}[1]{\mathrm{#1}}
\newcommand{\re}[0]{\mrm{Re}}
\newcommand{\im}[0]{\mrm{Im}}

\newcommand{\bea}{\begin{eqnarray}}
\newcommand{\eea}{\end{eqnarray}}

\newcommand{\nn}[0]{\nonumber}
\newcommand{\co}[0]{\mc{O}}
\newcommand{\gev}{{\rm GeV}}

\newcommand{\qq}{(q^2)}

\definecolor{Red}{rgb}{1.,0.,0.}

\definecolor{Green}{rgb}{0.2,.7,0.2}

\usepackage{hyperref}
\hypersetup{colorlinks,citecolor=nicegreen,linkcolor=nicered}
\hypersetup{colorlinks=true}

\begin{document}

\author{Damir Be\v cirevi\' c} \email[Electronic address:]{damir.becirevic@th.u-psud.fr}
\affiliation{Laboratoire de Physique Th\'eorique, CNRS/Universit\'e
  Paris-Sud 11 (UMR 8627), 91405 Orsay, France}

\author{Svjetlana Fajfer} \email[Electronic
address:]{svjetlana.fajfer@ijs.si} 
\affiliation{Department of Physics,
  University of Ljubljana, Jadranska 19, 1000 Ljubljana, Slovenia}
\affiliation{J. Stefan Institute, Jamova 39, P. O. Box 3000, 1001
  Ljubljana, Slovenia}

\author{Nejc Ko\v snik} 
\email[Electronic address:]{nejc.kosnik@ijs.si}
\affiliation{Department of Physics,
  University of Ljubljana, Jadranska 19, 1000 Ljubljana, Slovenia}
\affiliation{J. Stefan Institute, Jamova 39, P. O. Box 3000, 1001 Ljubljana, Slovenia}

\title{Lepton flavor non-universality in  $b \to s  \ell^+ \ell^-$ processes}

\date{\today}

\begin{abstract} 
  We explore a scenario of New Physics entering the description of
  $B\to K^{(\ast )} \mu \mu$ decay through couplings to the operators
  ${\cal O}_{9,10}^\prime$, satisfying $C_9^\prime = - C_{10}^\prime$.
  From the current data on ${\cal B}(B_s\to \mu\mu)$ and
  ${\cal B}(B\to K \mu\mu)_{[15,22] \gev^2}$, we obtain constraints on
  $\re C_{10}^\prime$ and $\im C_{10}^\prime$ which we then assume to
  be lepton specific, and find
  $R_K= {\cal B}(B\to K \mu\mu)/ {\cal B}(B\to K ee)_{[1,6]
    \gev^2}=0.88(8)$,
  consistent with recent value measured at LHCb. A specific
  realization of this scenario is the one with a scalar leptoquark
  state $\Delta$, in which $C_{10}^\prime$ is related to the mass of
  $\Delta$ and its Yukawa couplings. We then show that this scenario
  does not make any significant impact on $B_s-\overline B_s$ mixing
  amplitude nor to $ {\cal B}(B\to K\nu\bar \nu)$.  Instead, it can
  modify
  $R_{K^\ast}= {\cal B}(B\to K^\ast \mu\mu)/ {\cal B}(B\to K^\ast
  ee)_{[1,6] \gev^2}$,
  which will soon be experimentally measured and we find it to be
  $R_{K^\ast}= 1.11(8)$, while $R_{K^\ast}/R_K= 1.27(19)$. A similar
  ratio of forward-backward asymmetries also becomes lower than in the
  Standard Model.
\end{abstract}

\preprint{{\tt  LPT-Orsay-15-25}}

\maketitle

\section{Introduction}
The $b\to s$ transitions were in the focus of many theoretical and experimental studies during the last two decades due to the 
possibility to constrain potential New Physics (NP) contributions at low energies. With LHC7 and LHC8 runs direct searches for NP
became available.  This gives us an excellent opportunity to question
the appearance of physics beyond Standard Model (SM). At low energies $B$-factories and the LHCb experiment provided flavor physics community
with a lot of rather precise results on $b \to s$ transitions.  The
LHCb experiment has observed slight discrepancies between the SM predictions and the experimental results for the angular
observables in $B \to K^\ast \mu^+ \mu^-$ decay. This effect has been
attributed to NP, although the tension might be a result of
the SM QCD effects.  Recently, another anomaly in $b \to s \ell^+ \ell^-$
transition has been found in the ratio of the branching fractions,
\begin{equation}
\label{e1}
R_K = \frac{ \mc{B}( B \to K \mu^+ \mu^-)_{q^2 \in
    [1,6]\e{GeV^2}}}{\mc{B}( B \to K e^+ e^-)_{q^2 \in
    [1,6]\e{GeV^2}}} \,.
\end{equation}
LHCb Collaboration measured this ratio for the square of dilepton invariant mass in the bin $1$ GeV$^2  \leq  q^2 \leq 6 $   GeV$^2$, and found~\cite{Aaij:2014ora},  
\begin{equation}
\label{e2}
R_K^\mrm{LHCb} = 0.745 \pm^{0.090}_{0.074} \pm 0.036\,,
\end{equation}
lower than the SM prediction, $R_K^{SM} =1.0003 \pm 0.0001$, in which
next-to-next-to-leading QCD corrections have been included~\cite{Hiller:2003js}.
In other words, the LHCb result points towards a $2.6\ \sigma$ effect of the lepton flavor universality violation.

Furthermore, the combined data analysis of the $B_s \to \mu^+ \mu^-$
events gathered at LHCb and CMS resulted in
$\mc{B}(B_s \to \mu^+ \mu^-)=(2.8^{+0.7}_{-0.6}) \times
10^{-9}$~\cite{CMS:2014xfa},
in good agreement with the SM prediction
$\mc{B}(B_s \to \mu^+ \mu^-)=(3.65\pm 0.23) \times
10^{-9}$~\cite{Bobeth:2013uxa}.
This offers an excellent probe of $b \to s \mu^+ \mu^-$ transition in
the light of SM and gives rather tight constraints on parameter space
of many models of NP.  The $R_K$ anomaly has been approached in the
literature in different ways: either by using the effective Lagrangian
approach or in a specific model of NP.  For example the effective
Lagrangian approach used in
references~\cite{Alonso:2014csa,Hiller:2014yaa,Glashow:2014iga,Bhattacharya:2014wla}
indicated that in order to understand the measured value of $R_K$ one
must include the effects of NP, and that the effects of
non-perturbative QCD alone could not explain such a large deviation of
$R_K$ from
unity~\cite{Hiller:2014yaa,Kosnik:2012dj,Ghosh:2014awa,Biswas:2014gga,Hurth:2014vma,Glashow:2014iga,Altmannshofer:2014rta,Sahoo:2015wya,
  Varzielas:2015iva,Bhattacharya:2014wla,Gripaios:2014tna,
  Jager:2012uw,Descotes-Genon:2013wba,Altmannshofer:2013foa,
  Crivellin:2015mga, Sierra:2015fma}. In particular, it was found that the NP
contribution most likely affects $C_9 $, $C_{10}$ or $C_9^{\prime} $,
$C_{10}^{\prime}$ effective Wilson coefficients, and that some kind of
lepton flavor universality violation is needed, e.g.
$C_9^\mu \neq C_9^e$~\cite{Hiller:2014ula,Ghosh:2014awa}. In order to
determine whether $R_K$ anomaly is due to NP in electron or/and muon couplings
through a combined analysis of
several decay modes, it is very important to have a high
precision knowledge of hadronic form
factors~\cite{Gripaios:2014tna,Jager:2012uw,Descotes-Genon:2013wba},
which can be computed in the region of large $q^2$'s by means of
numerical simulations of QCD on the
lattice~\cite{Becirevic:2012fy,Bouchard:2013eph,Horgan:2013hoa}.

In this study we first use a model independent approach, assuming that
NP contributes at low energies to an operator that is a product of a
right-handed quark and a left-handed muon current. In the language of
$b \to s \mu \mu$ effective Hamiltonian such a situation corresponds
to a combination of Wilson coefficients $C_9^{\prime} $ and
$C_{10}^{\prime}$, and that they obey
$C_{9}^\prime = -C_{10}^{\prime}$. Decays to the final states with
electron-positron pair are instead governed by the SM only. This
assumption is motivated by the fact that measured quantities of
$b \to s e^+ e^-$ processes agree with the SM predictions better than
they do for the $b \to s \mu^+ \mu^-$ processes~\cite{Hurth:2014vma},
which are also more precisely measured than the electronic modes. We consider
simultaneously the constraints posed by
$ {\cal B}(B \to K \mu^+ \mu^-)$ and $ {\cal B}(B_s \to \mu^+ \mu^-)$
on such a scenario, and then predict the $R_K$ as well as
$R_{K^\ast}$. We discuss other observables which might serve as
additional probes of the observed lepton-flavor universality
violation.

A specific realization of the scenario we discuss in this paper is a
model with a light scalar leptoquark $\Delta$ with quantum numbers of
$SU(3)_c \times SU(2)_L \times U(1)_Y$ being $(3,2,1/6)$.  It indeed
verifies the relation,
$C_9^{\prime} = -C_{10}^{\prime}$~\cite{Kosnik:2012dj}, and leads to a
consistency with the measured value of $R_K$. The features of this
leptoquark state have been already described in the
literature~\cite{Dorsner:2014axa}. While there is no theoretical
motivation to forbid leptoquark contributing to $b \to s e e$ decays,
simultaneous presence of both muonic and electronic couplings could be
problematic because they would, together, induce lepton flavor
violation in $B_s \to e \mu$ and $\mu \to e \gamma$ decays.  It is
interesting that the flavor physics constraints at low energies agree
and are complementary with the constraints obtained from the direct
experimental searches at LHC~\cite{ATLAS:2012aq,CMS:zva}. Furthermore, the atomic parity
violation experiments provided a strong constraint on the interaction
of the down-quark--electron interaction with the leptoquark
state~\cite{Dorsner:2014axa,Gresham:2012wc}, while the couplings to
muons appear to be less constrained via
$ {\cal B}(K_L \to \mu^\pm e^\mp) < 4.7 \times
10^{-12}$~\cite{pdg,Dorsner:2014axa}.
We therefore assume in our analysis that in the
$b \to s \ell^+ \ell^-$ processes only the muons can interact with the
leptoquark state. A few other leptoquark states have been discussed in
the
literature~\cite{Hiller:2014yaa,Sahoo:2015wya,Kosnik:2012dj,Gripaios:2014tna}
as possible candidates to contribute to the $R_K$ anomaly. However,
the leptoquark with quantum numbers $(3,2,1/6)$ has a desired feature
that it can be light without destabilizing the
proton~\cite{Weinberg:1980bf,Buchmuller:1986iq,Dorsner:2012nq}. Notice also that another light
leptoquark scalar state, not mediating the proton decay, is
$(3,2,7/6)$ and it leads to the relation $C_9 =C_{10}$. That latter
scenario, however, cannot explain the $R_K$ anomaly as discussed in
Refs.~\cite{Hiller:2014yaa,Sahoo:2015wya}.

In Sec.~\ref{SEC2} we remind the reader of the main definitions and give  basic expressions for  $ {\cal B}(B_s \to \mu^+ \mu^-)$ and $ {\cal B}(B  \to K \mu^+ \mu^-)$, which are then used, together with the experimental data in Sec.~\ref{SEC3}, to 
 constraint $C_{10}^{\prime}=-C_9^{\prime}$ and show the consistency of our value for $R_K$ with the measured one at LHCb.  Furthermore, we make a prediction of the similar ratio in the case of $B\to K^\ast \ell^+\ell^-$ decays and discuss other observables that might be of interest for testing the lepton flavor universality violation. In Sec.~\ref{sec:lq} we discuss a model with scalar leptoquark in which the relation $C_{10}^{\prime}=-C_9^{\prime}$ holds exactly, and is connected to other similar processes involving the $b\to s$ transitions which we also discuss. We finally summarize our findings in Sec.~\ref{CONCL}.

\section{Effective Hamiltonian and basic formulas\label{SEC2}}
The processes with flavor structure $(\bar s b)\, (\bar \mu \mu)$ at scale
$\mu=\mu_b =4.8$~GeV are governed by dimension-6 effective Hamiltonian
~\cite{Grinstein:1988me,Misiak:1992bc,Buras:1994dj}:
\begin{equation}
  \label{eq:Heff}
    \mc H_\mrm{eff} = -\frac{4 G_F}{\sqrt{2}} V_{tb} V_{ts}^*
\left[\sum_{i=1}^6 C_i(\mu) \mc{O}_i(\mu) 
+ \sum_{i=7,\ldots,10} \left(C_i(\mu) \mc{O}_i(\mu) +C^\prime_i(\mu)
  \mc{O}^\prime_i(\mu) \right)  \right]\,.
\end{equation}
The contributions of the charged-current operators $\co_{1,2}$, QCD
penguins $\co_{3,\ldots,6}$, and the electromagnetic (chromomagnetic)
dipole operators $\co_7\,(\co_8)$ will be assumed to be saturated by
the SM. On the other hand, operators involving a quark and
a lepton current will contain the SM and potential NP 
contributions. The basis of operators may be further extended to
account for possible (pseudo)scalar or tensor
operators~\cite{Becirevic:2012fy}, whereas for the purposes of this
work the following operators will suffice:
\begin{equation}
  \label{eq:ops}
  \begin{split}
\co_{7} &= \frac{e}{g^2} m_b (\bar s \sigma_{\mu\nu} P_R b)
          F^{\mu\nu}\,,\qquad \co_{8}=\frac{1}{g} m_b (\bar s \sigma_{\mu\nu} G^{\mu\nu} P_R
         b)\,,\\
\mc{O}_9 &= \frac{e^2}{g^2} (\bar s \gamma_\mu P_L b) (\bar \ell
\gamma^\mu \ell)\,,\qquad \quad\! \mc{O}_{10} = \frac{e^2}{g^2} (\bar s \gamma_\mu P_L b) (\bar \ell
\gamma^\mu \gamma_5 \ell) \,.
  \end{split}
\end{equation}
Here $P_{L/R} = (1 \mp \gamma_5)/2$, while $e$ is the electromagnetic and
$g$ the color gauge coupling. $F^{\mu\nu}$ and $G^{\mu\nu}$ are the
electromagnetic and color field strength tensors, respectively.  The
basis is further extended by the wrong-chirality operators,
$\co'_{9,10}$, which are related to $\co_{9,10}$ by replacing 
$P_L \leftrightarrow P_R$  in the quark current.

\subsection{$B \to K \mu^+ \mu^-$}
In calculating the amplitude for the $B \to K \mu^+ \mu^-$ decay it is
convenient to group the combinations of  Wilson coefficients
multiplying the same hadronic matrix element.
Namely, the operators $\co_{1-6}$ mix at leading order into
$\co_{7,8,9}$ and it is customary to define effective Wilson
coefficients as~\cite{Buras:1993xp}:
\begin{equation}
\label{eq:Ceff}
\begin{split}
C_7^{\rm eff}(\mu_b) & =  \frac{4\pi}{\alpha_s}\, C_7 -\frac{1}{3}\, C_3 -
\frac{4}{9}\, C_4 - \frac{20}{3}\, C_5\, -\frac{80}{9}\,C_6\,,
\\
C_9^{\rm eff}(\mu_b) & =  \frac{4\pi}{\alpha_s}\,C_9 + Y(q^2)\,,
\\
C_{10}^{\rm eff} (\mu_b)& =  \frac{4\pi}{\alpha_s}\,C_{10}\,,\qquad
C_{7,8,9,10}^{\prime \mrm{eff}}(\mu_b) = \frac{4\pi}{\alpha_s}\,C^\prime_{7,8,9,10}\,,
\end{split}
\end{equation}
where the function $Y(q^2)$ at NLL can be found in
Ref.~\cite{Altmannshofer:2008dz}.  We also incorporate the NNLL mixing
of $\co_1$ and $\co_2$ into $\co_7$ and $\co_9$ as calculated in
Ref.~\cite{Greub:2008cy}. The Wilson coefficients on the right-hand
sides are evaluated at $\mu = \mu_b$. For the sake of readability we
will from here on discuss only the effective Wilson coefficients that
will be addressed simply as ``Wilson coefficients'' and denoted
without the ``eff'' label. The values of the SM Wilson coefficients at
scale $\mu_b$ are $C_7= -0.304$, $C_9 = 4.211$, and
$C_{10} =
-4.103$~\cite{Buras:1993xp,Bobeth:1999mk,Altmannshofer:2008dz}.

The decay spectrum as a function of the invariant mass of the muon
pair is given by
\begin{equation}
 \label{eq:GammaBK}
\frac{\mrm{d}\Gamma}{\mrm{d}q^2}(B \to K \mu^+ \mu^-)  = 2a_\mu (q^2) + \frac{2}{3} c_\mu(q^2)\,,
\end{equation}
where $q^2 = (p_{\mu^-} + p_{\mu^+})^2$, while functions $a_\mu(q^2)$,
$c_\mu(q^2)$ are combinations of Wilson coefficients and
hadronic form factors and their explicit expressions can be found in
Ref.~\cite{Becirevic:2012fy} and in the Appendix of the present paper in the limit of $m_\ell \to 0$. The rate depends on the sums of the
Wilson coefficients of opposite chiralities, $C_7 + C_7'$, $C_9 + C_9'$,
$C_{10} + C_{10}'$, from what follows that even in principle we cannot determine the
chirality of the quark-current in $B \to K \mu^+ \mu^-$.

Definitions of the hadronic form factors are relegated to the
Appendix. We employ the form factors calculated in the unquenched
lattice simulation using non-relativistic formulation of the $b$ quark
and staggered fermion formulation for the light
quarks~\cite{Bouchard:2013eph}.  We use the $z$-expansion to
parameterize the form factors and take into account the statistical
errors given by the covariance matrix of the parameters, both given
in~\cite{Bouchard:2013eph}. However, we neglect additional systematic
errors that should come on top of the ones contained in the covariance
matrix. The correlations between form factor parameters are propagated
onto observables of interest, namely we can construct $\chi^2$
statistic for $\mc{B}(B \to K \mu^+ \mu^-)$ and $R_K$, that are
functions of the form factor parameters, as well as the Wilson
coefficients. Nonlocal contributions to the decay amplitude due
  to operators $\mc{O}_{1,2}$ are taken into account by leading order in
  operator product expansion together with next-to-next-to-leading
  logarithmic QCD corrections~\cite{Greub:2008cy}. Higher orders in
  operator product expansion have been shown to have small
  effect in the large $q^2$ region~\cite{Grinstein:2004vb}.
Since the partial branching ratio that we are
  interested in corresponds to an integral over a large region of
  $q^2$ we rely on the semi-local quark-hadron duality~\cite{Beylich:2011aq}. In the SM
  limit the prediction of the branching ratio in the high-$q^2$ bin is
\begin{equation}
\label{eq:BK-SM}
\mc{B}(B^+ \to K^+\mu^+ \mu^-)|^\mrm{SM}_{q^2\in [15,22]\mrm{GeV}^2} =
(10.2 \pm 0.5) \E{-8}\,.
\end{equation}

The LHCb collaboration measured partial branching fractions below and
above the region of charmonium resonances. For the $q^2>15\e{GeV}^2$
region we can predict the partial branching ratio using form factors
determined on the lattice that are largely free from extrapolation
errors and parameterization dependence. Thus we will
use~\cite{Aaij:2014pli},
\begin{equation}
  \label{eq:brbkmm}
  \mc{B}(B^+ \to K^+\mu^+ \mu^-)|_{q^2\in [15,22]\mrm{GeV}^2} = (8.5
  \pm 0.3 \pm 0.4)\E{-8}\,,
\end{equation}
as an experimental constraint, where the errors quoted are statistical
and systematic, respectively. In our analysis we will sum the two 
and treat the observable with a Gaussian $\chi^2$.

\subsection{$B_s \to \mu^+ \mu^-$}
This decay receives contributions from operators with axial, scalar,
and pseudoscalar lepton currents, and, owing to the pseudoscalar nature
of the $B_s$ meson, the wrong-chirality Wilson coefficients will affect the decay with
opposite sign. In the absence of (pseudo)scalar operators, the
amplitude is proportional to the difference $C_{10} -
      C_{10}^{\prime}$:
\begin{align}
  \label{eq:bsmm}
  P = \frac{2m_\mu}{m_{B_s}} (C_{10} -
      C_{10}^{\prime}) \,,
\end{align}
and the ``theoretical'' branching ratio is expressed as
\begin{equation}
  \mc{B}(B_s \to \mu^+ \mu^-)^\mrm{th} = \mc{B}_0 |P|^2\,,\qquad
  \mc{B}_0  = \frac{f_{B_s}^2 m_{B_s}^3}{\Gamma_s} \frac{G_F^2 \alpha^2
  |V_{tb} V_{ts}|^2}{(4 \pi)^3} \,\sqrt{1-\frac{4 m_\mu^2}{m_{B_s}^2 }}\,.
\end{equation}
For the decay constant of the $B_s$ meson we take $f_{B_s} = (228 \pm 8)\e{MeV}$, consistent with the average made by FLAG~\cite{Aoki:2013ldr}.
Due to $B_s-\bar B_s$ oscillations and relatively large
$y_s = \Delta \Gamma_s/(2\Gamma_s)$ in the $B_s$ sector, the measured branching fraction
actually corresponds to a time-integrated rate of the oscillating
$B_s$ system to $\mu^+ \mu^-$~\cite{DeBruyn:2012wj}. In effect, the
value reported by the experimentalists is different from $\mc{B}(B_s
\to \mu^+ \mu^-)^\mrm{th}$:
\begin{align}
  \mc{B}(B_s \to \mu^+ \mu^-)^\mrm{exp} &=  \frac{\mc{B}_0}{1-y_s^2}  \left[|P|^2 +
                                          y_s \re (P^2) \right]\,.
\end{align}
Latest average of the LHCb and CMS measurements of $B_s \to
\mu^+ \mu^-$ branching fraction is~\cite{CMS:2014xfa}
\begin{equation}
  \label{eq:brbsmm}
  \mc{B}(B_s \to \mu^+ \mu^-)^\mrm{exp} = (2.8_{-0.6}^{+0.7})\E{-9}\,.
\end{equation}
The relative decay width difference $y_s = 0.061 \pm 0.009$ has been
determined from LHCb simultaneous measurement of total width
$\Gamma_s$ and width difference $\Delta \Gamma_s$
in decay channels $B_s \to J/\psi P^+ P^-$~\cite{Aaij:2014zsa}. The
above determined value agrees very well with the HFAG and PDG
averages~\cite{Amhis:2014hma,pdg}. In the fits we use the values for $\Gamma_s$ and $\Delta \Gamma_s$
reported by LHCb with summed statistical and systematic errors
\begin{equation}
   \Delta\Gamma_s = (0.0805 \pm 0.0123)\e{ps}^{-1}\,,\qquad  \Gamma_s =
   (0.6603 \pm 0.0042) \e{ps}^{-1}\,,
\end{equation}
with correlation coefficient $-0.3$~\cite{Aaij:2014zsa}.

\section{New Physics in $C_9' = -C_{10}'$ and prediction for $R_K$ \label{SEC3}}
We focus now on the SM extensions that affect the
effective Hamiltonian solely by a single operator that is a product of
right-handed quark current with a left-handed lepton current. In our
operator basis it corresponds to a linear combination
$\co_9' - \co_{10}'$ implying 
\begin{equation}
\label{eq:NP}
C_9'(\Lambda) = -C_{10}'(\Lambda)\,,
\end{equation}
where $\Lambda$ is a scale where NP degrees of freedom are integrated
out. An explicit example of such a scenario can be made in a
leptoquark model that will be discussed in Section~\ref{sec:lq}. If
Eq.~\eqref{eq:NP} holds at scale $\Lambda$ it is neccessary to run the
Wilson coefficients down to the low scale $\mu_b$ using the
renormalization group equations. Under QCD renormalization group the two operators do not run, keeping the
constraint~\eqref{eq:NP}
intact~\cite{Bobeth:2003at}.~\footnote{Eq.~(\ref{eq:NP}) is broken
  only by tiny effects from QED renormalization.}
Thus we have, at low energies, a SM modification that satisfies
\begin{equation}
C_9^{\prime} = -C_{10}^{\prime}\,,
\end{equation}
where $C_{9,10}^{\prime}$ are scale invariant, modulo small QED corrections.

\begin{figure}[!htbp]
  \centering
  \includegraphics[width=0.5\textwidth]{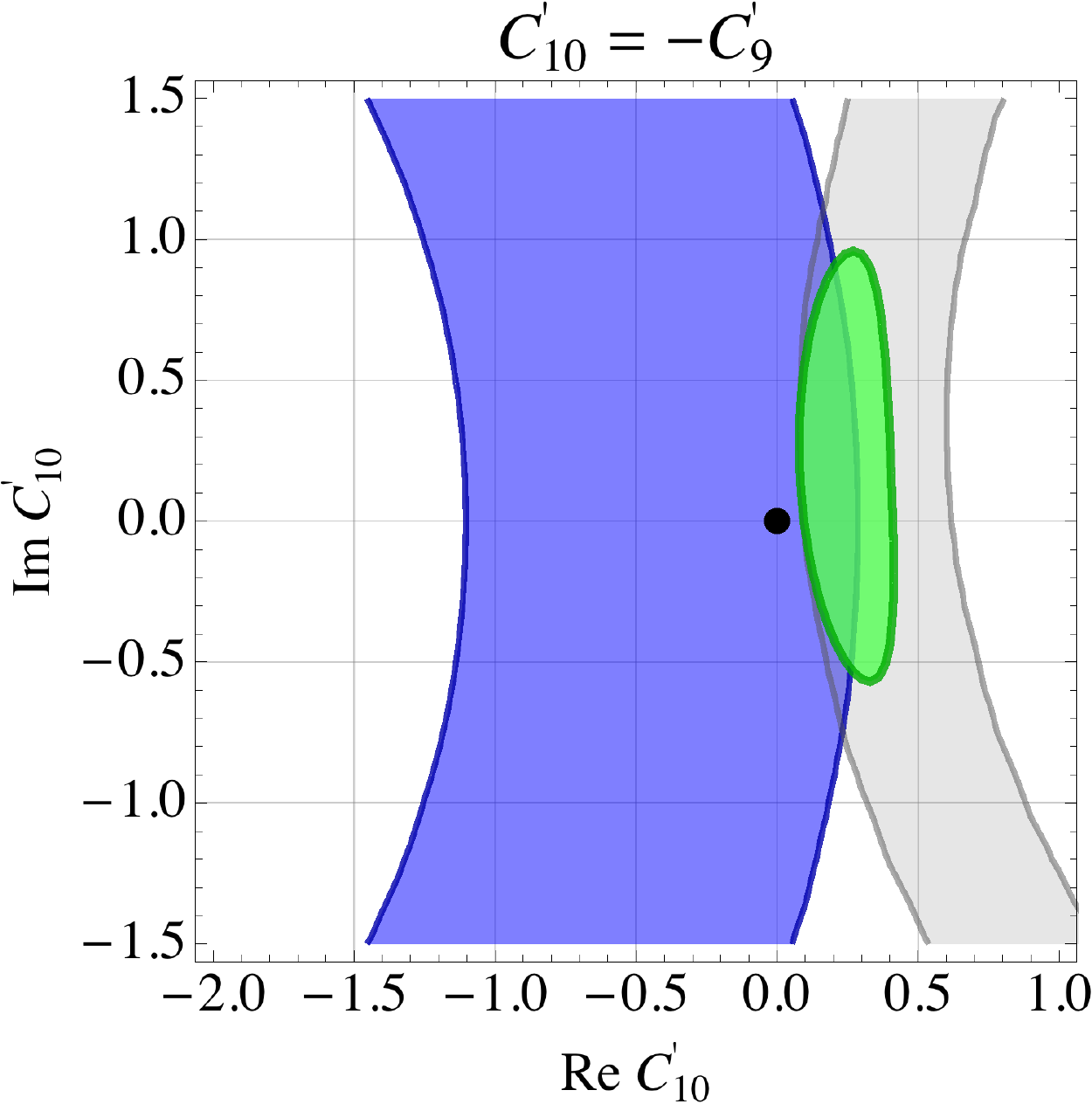}
  \caption{Regions in the complex $C_{10}^{\prime}$ plane
   that are in $1\sigma$ agreement with $B_s \to \mu^+ \mu^-$ (blue),
    $B \to K \mu^+ \mu^-$ (gray). Green area corresponds to the
   $1\sigma$ coverage of $R_K$ from fit to both observables. Black dot is the
   SM.}
  \label{fig:fit}
\end{figure}

In Fig.~\ref{fig:fit} we show in gray the $1\sigma$ region in the
$C_{10}^{\prime}$ plane as obtained from the fit to the 
partial branching fraction of $B^+ \to K^+ \mu^+ \mu^-$, cf. 
Eq.~\eqref{eq:brbkmm}. The $1\sigma$ region is defined here as $\chi^2 <
2.30$. The width of the ``donut'' reflects both experimental and form
factor uncertainties. The SM point in the parameter space is marked with a dot and
exhibits a tension with the measurement with $\chi^2 =
3.9$. 
In Fig.~\ref{fig:fit} the
$1\sigma$ region (defined as before) of fit to the $\mc{B}(B_s \to
\mu^+ \mu^-)$ according to Eq.~\eqref{eq:brbsmm} is depicted in blue. In this case the SM point is in comfortable
agreement with the observable ($\chi^2 = 0.7$). Then we perform combined fit to
all of the above quantities and find the best value to be $\chi^2_\mrm{min} =
2.26$, which is substantially better than the SM point with
$\chi^2_\mrm{SM} = 4.6$. The green patch is defined by $\chi^2 <
\chi^2_\mrm{min} + 1 = 3.26$ with $39\,\%$
  C.L. and corresponds to the $1\sigma$ region of predicted $R_K$
  given below.

\begin{figure}[!htbp]
  \centering
  \includegraphics[width=0.5\textwidth]{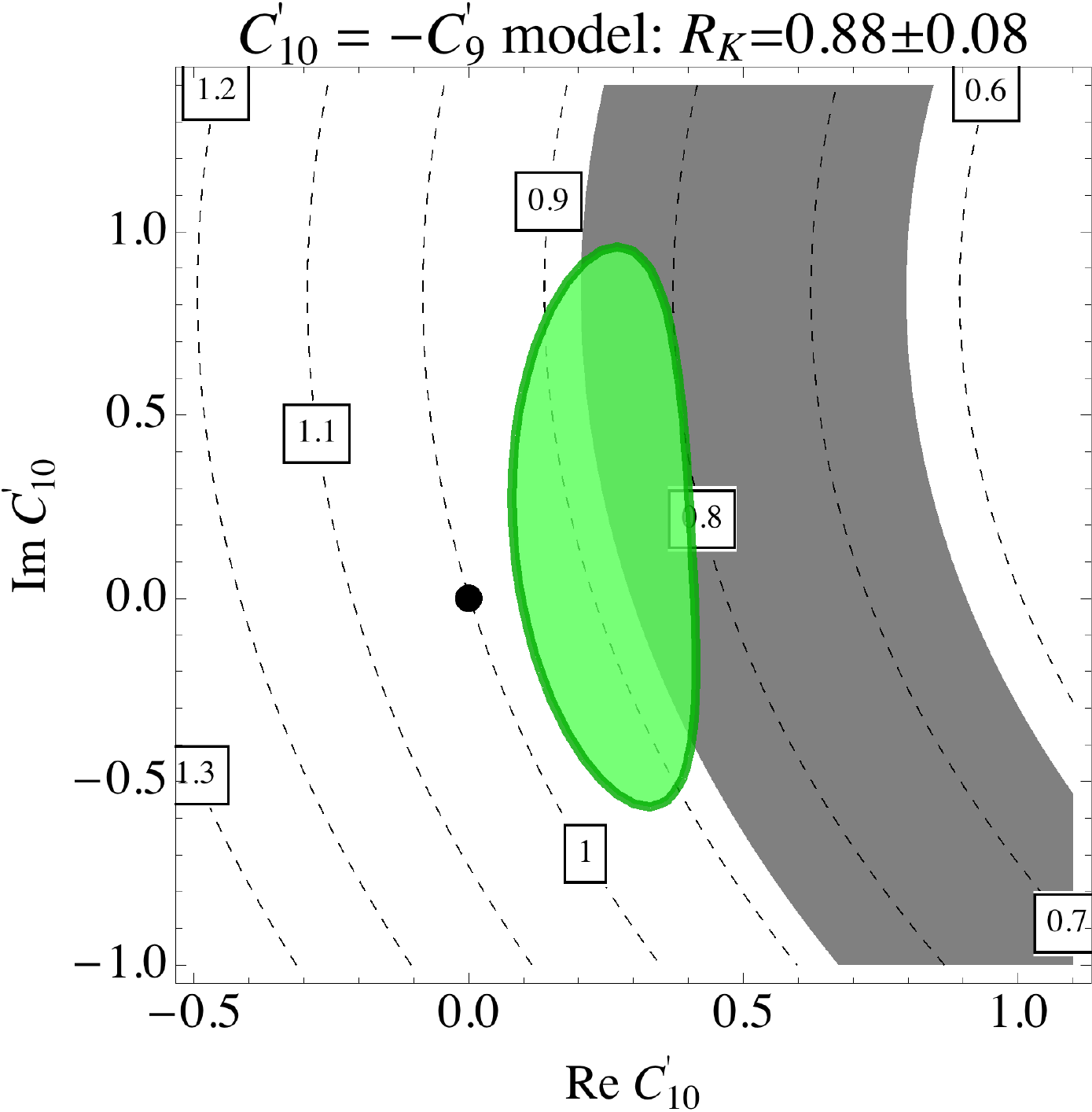}
  \caption{Contours of constant $R_K$ are indicated by dashed
    lines. Gray region represents the $1\sigma$ measured range of
    $R_K$ projected onto the $C_{10}^{\prime}$ plane, whereas green
    contour denotes the region allowed by $B_s \to \mu^+ \mu^-$ and
    $B \to K \mu^+ \mu^-$. Black dot is the SM.}
  \label{fig:RK}
\end{figure}

Assuming that the effective Hamiltonian \eqref{eq:Heff}, tailored for $b \to s e^+ e^-$,
receives only SM contributions, unlike $b \to s \mu^+ \mu^-$ that also receives NP contributions from $C_{9,10}^\prime$, we can now predict the value of $R_K$.
In $R_K$ the uncertainties of the hadronic form factors
cancel out to a large extent in the ratio and the formula boils down to:
\begin{equation}
\label{eq:rk-formula}
R_K (C_{10}^{\prime }) = 1.001(1) - 0.46\  \re [C_{10}^{\prime}] - 0.094(3)\  \im [C_{10}^{\prime}] +
0.057(1) |C_{10}^{\prime}|^2\,.
\end{equation}
Remaining uncertainties are indicated by the numbers in parentheses.
In Fig.~\ref{fig:RK} we show contours of constant $R_K$ in the $C_{10}'$ plane 
using the formula~\eqref{eq:rk-formula} with central values for the
coefficients. By dark gray we indicate the region corresponding  to 
the measured value of $R_K$. In the same figure we plot again 
the $1\sigma$ prediction of $C_{10}'$, also shown in Fig.~\ref{fig:fit}. We see an appreciable
overlap with the measured $R_K$. Mapping the fitted region (green) to
$R_K$ we obtain the prediction
\begin{equation}
  \label{eq:RKpred}
  R_K^\mrm{pred.} = 0.88 \pm 0.08\,,
\end{equation}
which is indeed in good agreement with $R_K^\mrm{LHCb} = 0.745 \pm^{0.090}_{0.074} \pm 0.036$~\cite{Aaij:2014ora}.

\subsection{Impact on $B\to K^\ast \ell^+ \ell^-$}

$B\to K^\ast \ell^+\ell^-$ is particularly interesting for the NP searches because of the observables that one can construct from the $q^2$-dependent coefficients $I_{1\dots 9}$ which appear in the angular distribution, 
\begin{align}\label{distr-1}
{d^4\Gamma(\bar B^0 \to   \bar K^{\ast 0} \ell^+\ell^-)\over dq^2\ d\cos\theta_\ell\ d\cos\theta_K\  d\phi} &=  {9\over 32\pi} \Biggl[ I_1^s \sin^2 \theta_K + I_1^c \cos^2\theta_K + (I_2^s  \sin^2 \theta_K + I_2^c \cos^2\theta_K) \cos 2\theta_\ell \Biggr. \nn\\
&+ I_3 \sin^2 \theta_K \sin^2 \theta_\ell \cos 2\phi +  I_4  \sin 2 \theta_K \sin 2 \theta_\ell \cos \phi + I_5  \sin 2 \theta_K\sin\theta_\ell\cos\phi \nn\\
&+(I_6^s \sin^2\theta_K+I_6^c \cos^2\theta_K)\cos\theta_\ell +I_7  \sin 2\theta_K\sin\theta_\ell\sin\phi\\
&\Biggl. +I_8 \sin 2\theta_K\sin 2\theta_\ell\sin\phi + I_9 \sin^2\theta_K\sin^2\theta_\ell \sin 2\phi\, \Biggr] .\nn
\end{align}
Differential decay rate is then simply $d\Gamma/dq^2 = (3 I_1^c + 6 I_1^s - I_2^c - 2 I_2^s)/4$, and the similar expressions can be written for the transverse/longitudinal part of the decay rate, for the forward-backward asymmetry, $A_{\rm fb}(q^2) = 3 I_6^s/(4 d\Gamma/dq^2)$,
CP-asymmetry, and several other observables. Each of the coefficient functions, $I_i\equiv I_{i}(q^2)$,  can be written in terms of transversity amplitudes, $A_{\perp,\parallel,0}^{L,R}(q^2)$, which are related to the respective spin states of the {\it on-shell} $K^\ast$-meson, and 
the amplitude $A_{t}^{L,R}(q^2)$ which is related to the {\it off-shell} virtual gauge boson decaying into the lepton pair. The superscripts $L,R$ indicate the chirality of the lepton. Detailed expressions can be found, for example, in Refs.~\cite{Altmannshofer:2008dz,Bobeth:2008ij,Descotes-Genon:2013vna,Becirevic:2011bp}.  

The strategy of looking for the NP effects through a detailed analysis of the angular distribution of  $B\to K^\ast \ell^+\ell^-$ is somewhat plagued by hadronic uncertainties. The observables built up of $A_{\perp,\parallel}^{L,R}(q^2)$ turn out to be 
less sensitive to hadronic uncertainties because they involve the (combinations of) hadronic form factors which appear to be under a rather good theoretical control, especially in the region of small $q^2$'s~\cite{Charles:1998dr,Beneke:2001at,Beneke:2004dp} (see also discussion in Ref.~\cite{Becirevic:2011bp}). On the other hand, the observables made of $A_{0,t}^{L,R}(q^2)$ 
entail the hadronic form factors that are less well understood. Moreover, the latter observables are subject to another kind of hadronic uncertainty, i.e. the one arising from misidentification of the $K\pi$ pairs coming from $B\to K^\ast (\to K\pi) \ell^+\ell^-$ with those emerging from $B\to K_0^\ast (\to K\pi) \ell^+\ell^-$, where $K_0^\ast$ stands for a broad scalar state~\cite{Becirevic:2012dp,Das:2014sra,Blake:2012mb,Matias:2012qz,Doring:2013wka}. Finally, and to avoid problems of the $c\bar c$-resonances in the $q^2$-spectrum of the decay,  a standard strategy is to either work at low $q^2 < m_{J/\psi}^2$ or large $q^2 \gtrsim 15 \gev^2$, in which the impact of the $c\bar c$-resonances is expected to be small. To be more specific, we fully rely on quark-hadron duality since we avoid the region in which the prominent narrow $c\bar c$-resonances appear, and integrate over a window $\gtrsim 5~\gev^2$.~\footnote{For a recent attempt to more realistically model the effects of such resonances see Ref.~\cite{Lyon:2014hpa} or those discussed previously in Refs.~\cite{Ali:1991is,Kruger:1996dt}. }

\begin{figure}[!htbp]
  \centering
  \includegraphics[width=0.45\textwidth]{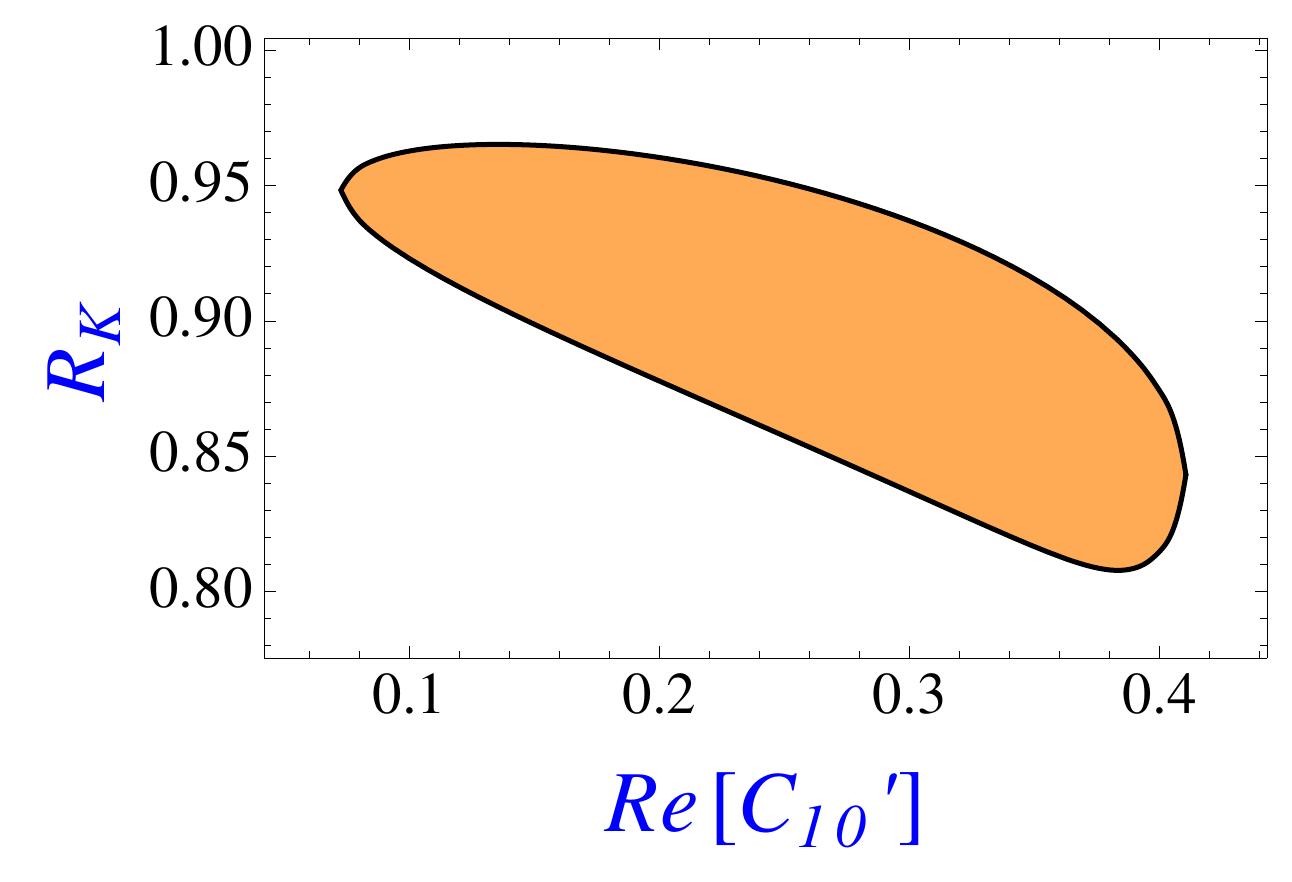}~ \includegraphics[width=0.45\textwidth]{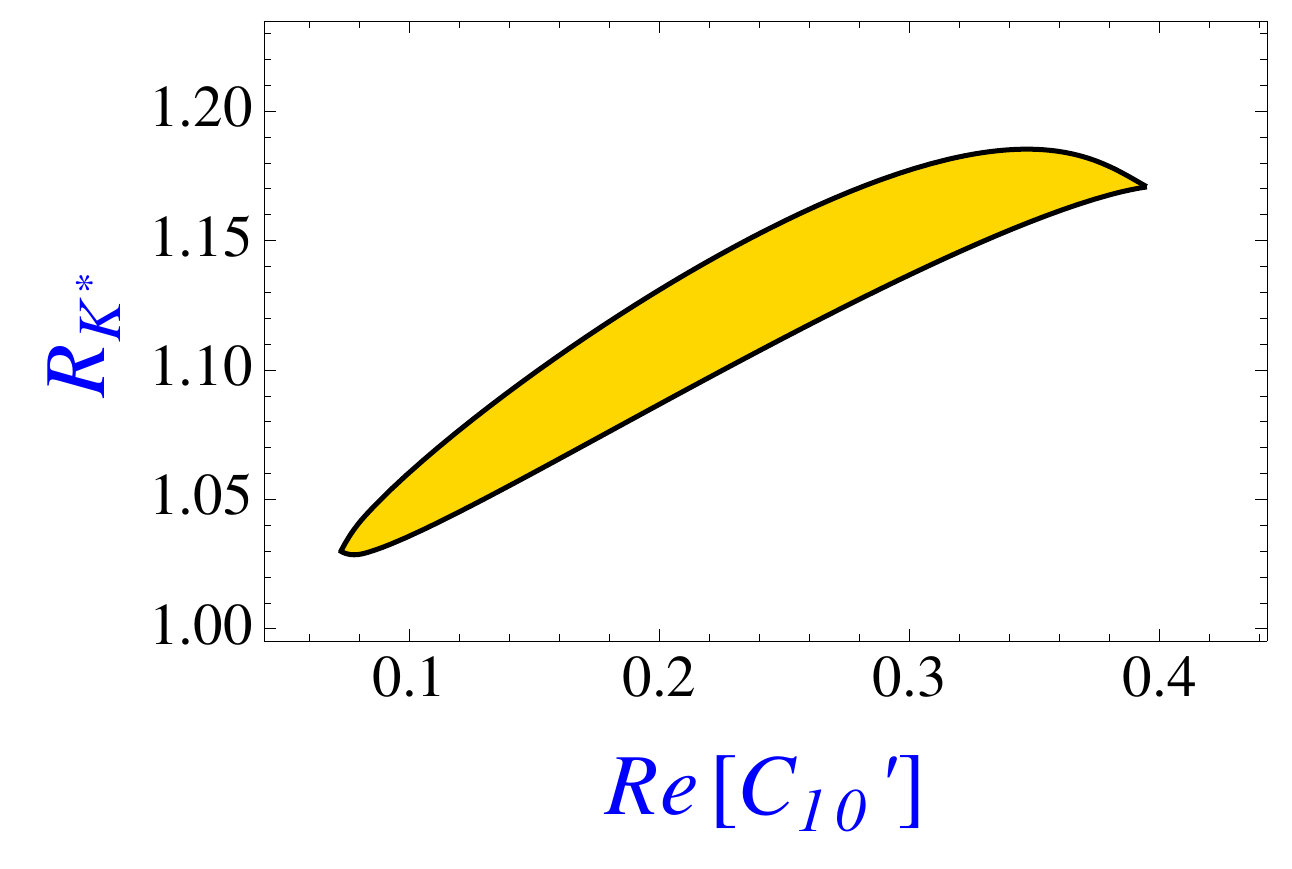}\\
  \includegraphics[width=0.45\textwidth]{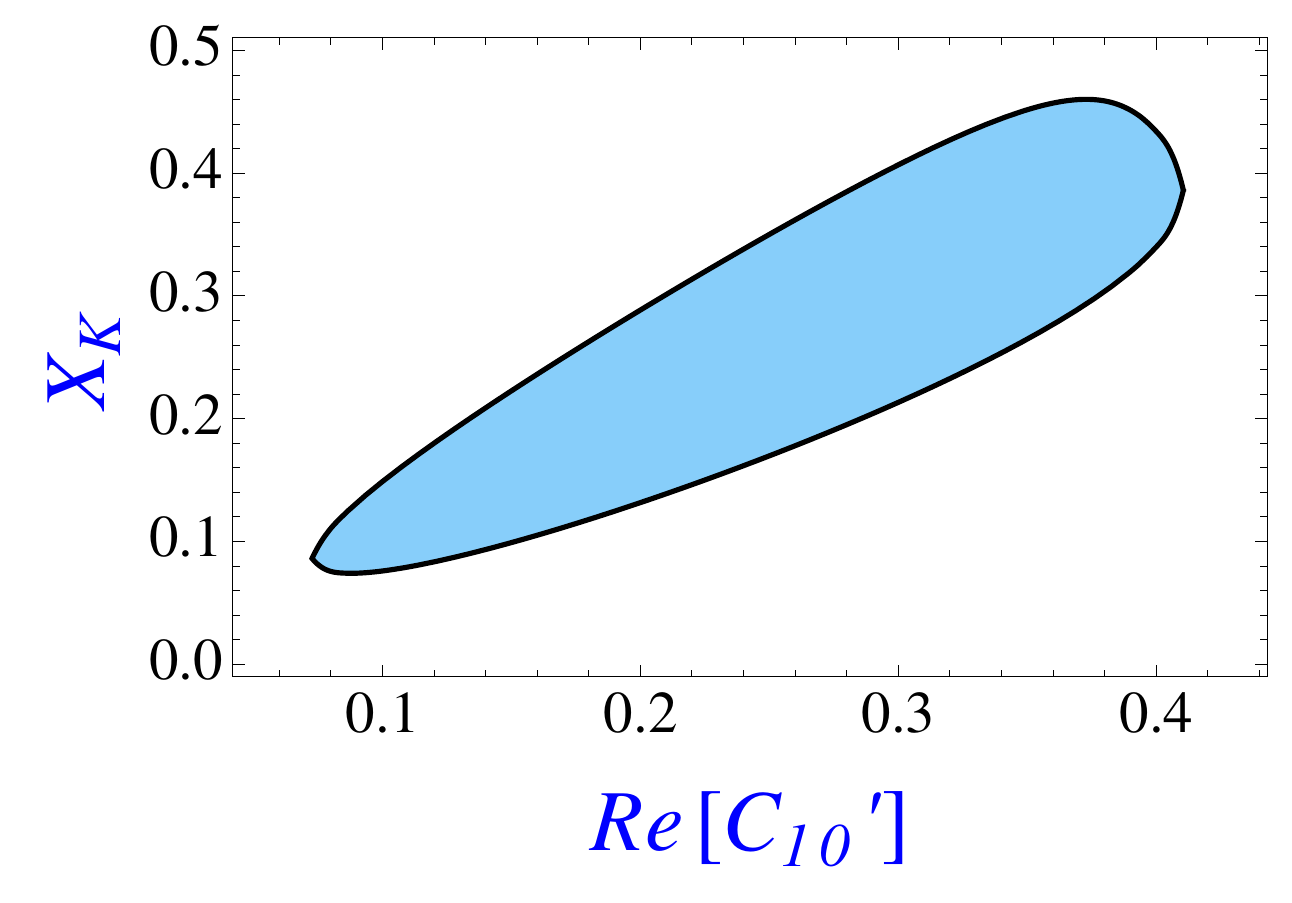}~ \includegraphics[width=0.45\textwidth]{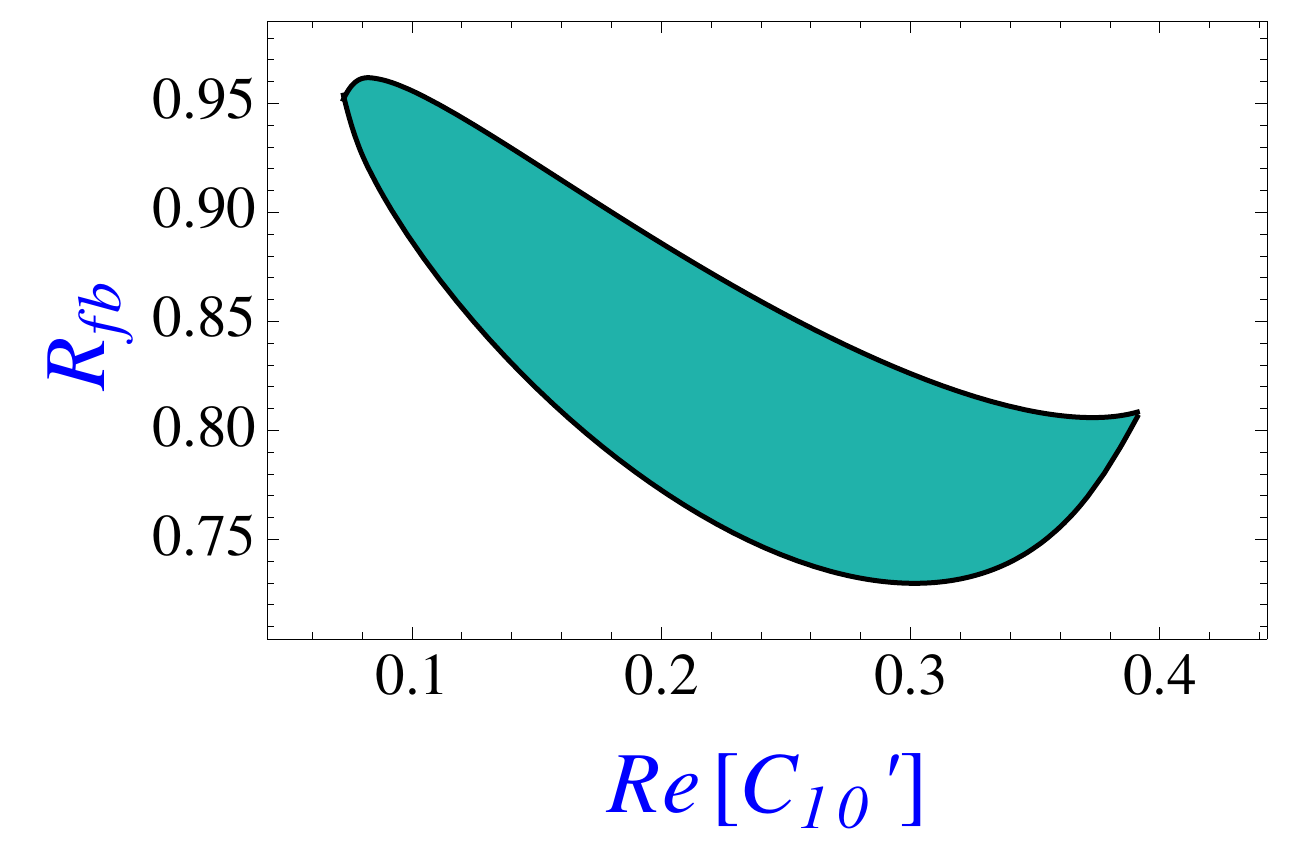}
  \caption{$R_K$, $R_{K^\ast}$, $X_K$ and $R_{\rm fb}$, defined in Eq.~(\ref{e1},\ref{def:RKst},\ref{def:XK},\ref{def:Rfb}) respectively, are plotted as functions of ${\rm Re}[C_{10}^\prime]$, in the range allowed by the measured values of ${\rm B}(B_s\to \mu^+\mu^- )$ and 
$\mc{B}(B\to K\mu^+ \mu^- )_{q^2 > 15\ \gev^2}$.  Instead of a curve for
each quantity we actually have a region of values, reflecting the fact
that for each ${\rm Re}[C_{10}^{\prime }]$ there is a range
of allowed values of ${\rm Im}[C_{10}^{\prime }]$, as shown in Fig.~\ref{fig:fit}.}
  \label{fig:Kstar}
\end{figure}

With the information obtained in the previous section of this paper,
i.e. with $C_{10}^{\prime } = -C_{9}^{\prime}$ extracted from the
comparison of the measured $\mc{B}(B_s\to \mu^+ \mu^-)$ and
$\mc{B}(B\to K\mu^+ \mu^-)_{q^2>15\ \gev^2}$ with the corresponding
theoretical expressions, we already showed that we were able to verify
the consistency of our result for $R_K$ with the one measured at
LHCb. With our approach, in which only the decay to muon-pair is
modified, we can also predict
$R_{K^\ast}$, defined as \bea\label{def:RKst} R_{K^\ast} = {
  \Gamma(B\to K^\ast \mu^+ \mu^- )_{q^2\in [1, 6]\ \gev^2}\over
  \Gamma(B\to K^\ast e^+e^-)_{q^2\in [1, 6]\ \gev^2} }, \eea as well
as the ratio of the two~\cite{Hiller:2014yaa,Hiller:2014ula}, namely,
\bea
\label{def:XK} X_K = {R_{K^\ast}\over R_K}-1\,.  
\eea 
In Ref.~\cite{Altmannshofer:2013foa} it was shown that the ratio of
forward-backward asymmetries integrated between $q^2\in [4,6]~\gev^2$
can also be sensitive to lepton flavor universality violation. After
defining, 
\bea A_{\rm fb [4-6]}^{\ell }= \frac{3}{4} \,
{\displaystyle{\int_{4\ \gev^2}^{6\ \gev^2} I_6^s(q^2)\ dq^2} \over
  \quad \Gamma(B\to K^\ast \ell^+ \ell^- )_{q^2\in [4, 6]\ \gev^2}
}\,, \eea the ratio of forward-backward asymmetries is then simply,
\bea \label{def:Rfb} R_{\rm fb} = {A_{\rm fb [4-6]}^\mu \over A_{\rm
    fb [4-6]}^e }\,.  
\eea 
To compute the above-mentioned quantities
we use the standard values of the Wilson
coefficients~\cite{Bobeth:1999mk}, and include the effect of quark
loops in the coefficients $C_{7,9}$ arising from the operators
$O_{1,2}$, as calculated in Ref.~\cite{Greub:2008cy}. We neglect the
soft gluon corrections to the charm quark loop at low $q^2$, which
according to Ref.~\cite{Khodjamirian:2010vf} is reasonable. At low
$q^2$ the hard scattering contributions are neglected.
For the
form factors we use the values computed by means of QCD sum rules on
the light cone~\cite{Ball:2004rg}.
In Fig.~\ref{fig:Kstar} we show our
results for $R_{K^\ast}$, $X_K$ and $R_{\rm fb}$ as functions of
Re[$C_{10}^{\prime }$]. For an easier comparison, in the same plot we
also show $R_K$.  The range
$0.075 \leq {\rm Re}[C_{10}^{\prime }] \leq 0.41$ has been obtained in
the previous section of this paper, where we showed that for a given value of
$0.075 \leq {\rm Re}[C_{10}^{\prime}] \leq 0.41$ there is a region of
allowed Im[$C_{10}^{\prime }$], and therefore instead of curves in
Fig.~\ref{fig:Kstar} we actually have the corresponding regions of
values determined by ${\rm Im}[C_{10}^{\prime }]$. We should emphasize
again that the uncertainties related to form factors cancel to a large
extent in the ratios.
As for the results, we first see that in the scenario with $C_{10}^{\prime} = - C_{9}^{\prime} \neq 0$, allowing coupling to muons only, and explicitly realized in the model with a $(3,2,1/6)$  leptoquark state, we get 
\begin{equation}
\label{eq:resKst}
\begin{split}
R_K  &= 0.88\pm 0.08\,, \qquad R_{K^\ast}= 1.11\pm 0.08\,,\\
X_K  &= 0.27\pm 0.19\,, \qquad R_{\rm fb} = 0.84\pm 0.12\,,  
\end{split}
\end{equation}
which are obviously different from the values obtained in the SM, $R_K^{\rm SM}=1.00$, $R_{K^\ast}^{\rm SM}=0.996(5)\approx 1$, $R_{\rm fb} = 0.995(4) \approx 1$, and $X_K=-0.004(5) \approx 0$.  
Notice, however, that while our value for $R_K$ is lower than the one in the SM, our prediction for $R_{K^\ast}$ is larger than that obtained in the SM. 
The measurement of $R_{K^\ast}$ at LHCb will therefore help to either confirm or discard our model as a viable description of the lepton flavor universality violation. The errors in Eq.~(\ref{eq:resKst}) are completely dominated by the range of $\re [C_{10}^\prime]$ and $\im [C_{10}^\prime]$, while those arising from form factors are reduced in the ratios and induce an uncertainty negligible in comparison with that coming from the variation of $C_{10}^\prime$.

Besides the above quantities, one can also check on the asymmetries $A_T^{(2)}$ and 
$A_T^{ ( \mathrm{Im})}$, defined in the Appendix, which are experimentally more difficult to study 
but which could be very useful to compare with predictions as their values can considerably change if $C_{9,10}^\prime \neq 0$. 
To exemplify that feature we consider the bin $q^2\in [2,6]~\gev^2$ and in the SM we have  
$\langle A_T^{(2)}(q^2)\rangle_{q^2\in [2,6]\ \gev^2}^{\mathrm{SM}} =-0.05(2)$, 
and $\langle A_T^{({\rm Im})}(q^2)\rangle_{q^2\in [2,6]\ \gev^2}^{\mathrm{SM}} \simeq 0$, for either 
electrons or muons in the final state. If, instead, the coefficients $C_{10}^\prime = - C_9^\prime$ 
become non-zero and take values within the green region shown in Fig.~\ref{fig:fit}, then in the case of $B\to K^\ast \mu^+\mu^-$, the above values change to 
\bea
\langle A_T^{(2)}(q^2)\rangle_{q^2\in [2,6]\ \gev^2}^\mu \simeq \biggl. -0.09\biggr|_{\mathrm{for }\ \re[C_{10}^\prime] = 0.08}, 
-0.24\ \biggr|_{\mathrm{for }\ \re[C_{10}^\prime] = 0.4},\qquad 
|\langle A_T^{({\rm Im})}(q^2)\rangle_{q^2\in [2,6]\ \gev^2}^\mu | \lesssim 0.27\,.
\eea
Notice also that $\langle A_T^{({\rm Im})}(q^2)\rangle$ has not yet been measured, and that the current errors on $\langle A_T^{(2)}(q^2)\rangle$ are 
still too large for making a meaningful quantitative comparison with our results~\cite{Aaij:2013iag}.

Finally, before closing this part of our paper, we need to comment on $P_5^\prime (q^2)$, an observable constructed from coefficients of the angular distribution of the $B\to K^\ast \ell^+\ell^-$ decay~\cite{DescotesGenon:2012zf}, 
$P_5^\prime (q^2)= I_5/\sqrt{-4 I_2^c I_2^s}$, which has been measured at LHCb, and turned out to be $4\sigma$ away from the value predicted in the SM when integrated over an interval $q^2\in [4.3,8.68]~\gev^2$~\cite{Aaij:2013qta}. 
More specifically, the SM value is $\langle P_5^\prime \rangle_{[4.3 - 8.68]}^{\rm SM}= - 0.90(5)$, while the measured one is $\langle P_5^\prime \rangle_{[4.3 - 8.68]}^{\rm LHCb}= - 0.19(16)$~\cite{Aaij:2013qta}, which can be
compactly written as, $\langle P_5^\prime \rangle_{[4.3 - 8.68]}^{\rm LHCb}/\langle P_5^\prime \rangle_{[4.3 - 8.68]}^{\rm SM} =0.22(18)$.  While the interpretation of this discrepancy is somewhat controversial~\cite{Descotes-Genon:2013wba,Altmannshofer:2013foa,Jager:2014rwa}, it is nevertheless interesting to check whether or not the leptoquark model used in this paper (and discussed in more details in the following Section) can describe the manifest disagreement between theory and experiment.  
With the values of $C_{10}^{\prime } = - C_9^{\prime }$ discussed above we indeed see that $\langle P_5^\prime \rangle_{[4.3 - 8.68]}^{\rm LQ}/\langle P_5^\prime \rangle_{[4.3 - 8.68]}^{\rm SM}   <1$, but with the leptoquark model discussed here 
we cannot reach very low values. We instead obtain  $0.78\leq \langle P_5^\prime \rangle_{[4.3 - 8.68]}^{\rm LQ}/\langle P_5^\prime \rangle_{[4.3 - 8.68]}^{\rm SM} \leq 0.98$. A similar tendency is observed for other bins, and in particular the one 
corresponding to $q^2\in [1,6]~\gev^2$.

\section{Model with a Scalar Leptoquark \label{sec:lq}}

In this Section we discuss a specific model in which the scenario discussed above, i.e. $C_9^\prime =-C_{10}^\prime$, is explicitly realized and involves the presence of a light scalar leptoquark state $\Delta$. 
More specifically, we choose the leptoquark $\Delta$ to carry the quantum numbers $(3,2,1/6)$ of the SM gauge group. Its couplings to fermions are described by a renormalizable Lagrangian
\begin{equation}
 \label{eq:Lag16}  
\begin{split}
  \mc{L} &= Y_{ij} \overline L_i \, i\tau^2 \Delta^* d_{Rj} 
+ \mrm{h.c.} \\
&= Y_{ij} \left(-\bar \ell_{Li} d_{Rj} \Delta^{(2/3)*} 
+ \bar \nu_{Lk}(V^\mrm{PMNS})^\dagger_{ki} d_{Rj} \Delta^{(-1/3)*} 
\right) + \mrm{h.c.}\,, 
\end{split}
\end{equation}
where $Y$ is a $3\times 3$ complex matrix, $L_i$ and $d_{Rj}$
are the lepton doublet and down-quark singlet. Charge eigenstates of the leptoquark doublet are denoted with
$\Delta^{(2/3)}$ and $\Delta^{(-1/3)}$ and we will assume that they
are degenerate. The second line in the above
Lagrangian is written in the fermion mass basis, and a relative
PMNS rotation in lepton doublet components has been assigned to the
neutrino sector.

Clearly, the lepton flavor universality is explicitly broken by the terms presented in Eq.~(\ref{eq:Lag16}). 
This might appear questionable because in a similar situation in which the coupling of
leptoquark to $\mu c$ would be allowed, the ratio of the electronic and
muonic widths of the decay of $J/\psi$ and its radial excitations have
been accurately measured, and shows no violation of the lepton flavor
coupling universality. In particular, the measured
$\Gamma(J/\psi\to \mu^+ \mu^- )/\Gamma(J/\psi\to e^+ e^- ) = 1.0016 \pm
0.0031$~\cite{pdg}
is in excellent agreement with its SM value,
$1.00001$.~\footnote{By explicitly including the lepton mass in the
  calculation of phase space we obtain, \bea \Gamma(J/\psi \to
  \ell^+\ell^-) = {16\pi \alpha^2\over 27 m_{J/\psi}} \left( 1 + {2
      m_\ell^2 \over m_{J/\psi}^2}\right) \sqrt{ 1 - {4 m_\ell^2 \over
        m_{J/\psi}^2} } \ f_{J/\psi}^2 \nn \eea and the effect on the
    ratio of the electronic and muonic widths is extremely small.}
  That situation is, however, much different from the examples
  discussed in this paper, because the amplitude for
  $J/\psi\to \ell^+\ell^-$ is dominated by the tree-level electromagnetic
  interaction diagram which is much larger than the weak interaction
  one, suppressed by $1/m_Z^2$ with respect to the dominant one, and
  therefore completely negligible. Our leptoquark state is
  $m_\Delta \gg m_Z$, and its contribution to $J/\psi\to \ell^+ \ell^- $ is
  even smaller than the weak interaction diagram and cannot make an
  impact on the decay of charmonia at the present level of accuracy.

  Instead, the weak $b\to s\mu^+\mu^-$ decays in the SM are
  loop-induced so that the tree level contribution involving couplings
  to the leptoquark state may become comparable in size to the
  SM amplitude, which is why the $b\to s\mu^+ \mu^- $ is
  likely to be more sensitive to the presence of the term described by
  the lagrangian~(\ref{eq:Lag16}). The relevant leptoquark coupling for the
  $b\to s\mu^+\mu^-$ is the product $Y_{\mu b} Y_{\mu s}$, which enters
  the Wilson coefficients divided by $m_\Delta^2$. The scalar particle
  exchange generates scalar operators in the Fierzed basis and those
  appear as (pseudo)vector currents in the ordinary operator basis~\cite{Kosnik:2012dj}:
\begin{equation}
\label{C10LQ}
C_{10}^{\prime} = -C_{9}^{\prime}  = \frac{\pi}{2\sqrt{2} G_F
  V_{tb} V_{ts}^* \alpha}\, \frac{Y_{\mu b} Y_{\mu s}^*}{ m_\Delta^2}\,.
\end{equation}
 We assume other elements of Yukawa matrix $Y$ to vanish.
The same state will also contribute at loop level to electro- and
chromo-magnetic operators $C_7^\prime(m_\Delta)$ and $C_8^\prime(m_\Delta)$ where
these coefficients will be suppressed by electromagnetic $\alpha(m_\Delta)/(4\pi)$ and
strong $\alpha_S(m_\Delta)/(4\pi)$ couplings at high scale $m_\Delta$,
respectively.
We have explicitly checked that these modifications result in
negligibly small value of $C_7^{\prime}$ when compared to
the $C_7$ of SM, cf. Eq.~\eqref{eq:Ceff}.
In the remainder of this Section we will analyze additional observables that constrain
this leptoquark scenario.

The considered leptoquark state $\Delta$ couples to the neutrinos
with the same couplings as to the charged leptons, only modified by a PMNS
rotation matrix. Namely, the charge $-1/3$ state will generate $(\bar
s b) (\bar \nu \nu)$ operators while the box diagrams will lead to
$B_s-\bar B_s$ mixing.

\subsection{Contribution of $(3,2,1/6)$  leptoquark in $B_s - \bar B_s $ oscillation frequency}
The state $(3,2,1/6)$ will induce $\Delta B=2$ box diagrams with $\mu$
and $\Delta^{(2/3)}$ or $\nu$ and $\Delta^{(-1/3)}$ running in the
box. The two contributions of boxes with $\mu$ and $\nu$ are equal in
the $m_\mu = 0$ limit and in sum they
amount to
\begin{equation}
\label{box-c}
  C_6^\mrm{LQ} (m_\Delta)= -\frac{Y_{\mu b}^{*2} Y_{\mu s}^2}{64 \pi^2  m_\Delta^2} \,.
\end{equation}
The effective $\Delta B = 2$ Hamiltonian is defined as 
\begin{equation}
\label{eq:Bsham}
{\cal H}_\mrm{eff} =
C_1^{\rm SM} (\bar b \gamma_\mu P_L s)\, (\bar b \gamma^\mu P_L
s) +  C_6^{\rm LQ}  (\bar b \gamma_\mu P_R s)\, (\bar b \gamma^\mu P_R
s)  +\mrm{h.c.}\,,
\end{equation}
where $P_{L/R} = (1\mp\gamma_5)/2$.  The coefficient in Eq.~\eqref{box-c} is
subject to QCD renormalization and has to be evaluated at scale
$\mu_b$. The anomalous dimensions of $C_6^{\rm LQ}$ is however
equal to the one of $C_1^{\rm SM}$. Therefore the two Wilson
coefficients renormalize with the same multiplicative factor between
scales $\mu = m_t$, where SM is matched onto effective Hamiltonian~\eqref{eq:Bsham}, and
$\mu_b$, where the hadronic matrix elements are computed. Remaining
$C_6^\mrm{LQ}$ running from $m_\Delta$ down to $m_t$ is already in the
asymptotic regime of QCD and can be safely neglected.
The mass difference of the $B_s - \bar B_s $ system is then
\bea
\Delta m_{B_s} =\frac{2}{ 2 m_{B_s}  }\left| \frac{G_F^2
    m_W^2}{16\pi^2} (V_{tb}^\ast V_{ts} )^2 \eta_B S_0(x_t) + \frac{\eta_B}{4} C_6^{\rm LQ}(m_\Delta) \right| 
\langle \bar B_s^0\vert \bar b \gamma_\mu (1-\gamma_5) s\, \bar b \gamma^\mu (1-\gamma_5) s \vert B_s^0\rangle\,.
\eea
By using Eq.~(\ref{C10LQ})  we can write
\bea
C_6^\mrm{LQ}(m_\Delta)= - {G_F^2 \over 8 \pi^4 } (V_{tb}^\ast V_{ts} )^2 \alpha^2
m_\Delta^2 (C_{10}^{\prime \ast})^2\,,
\eea 
which, together with $\langle \bar B_s^0\vert \bar b \gamma_\mu (1-\gamma_5) s\, \bar b \gamma^\mu (1-\gamma_5) s \vert B_s^0\rangle = (8/3) f_{B_s}^2 m_{B_s}^2 B_{B_s}$, gives 
\bea \label{eq:dm}
\Delta m_{B_s} = \underbrace{ {G_F^2 m_W^2\over 6\pi^2 }
  |V_{tb}^\ast V_{ts} |^2  f_{B_s}^2 m_{B_s} B_{B_s}  \eta_B S_0(x_t)
}_{\Delta m_{B_s}^{\rm SM}} \left| 1 -\frac{1}{2\pi^2}  {  \alpha^2 \over
  S_0(x_t) }  (C_{10}^{\prime \ast})^2 \frac{m_\Delta^2}{m_W^2}\right|\,.
\eea
With  the current values for $f_{B_s}= 228(8)$~MeV and $B_{B_s}=1.33(6)$, as obtained in numerical simulations of QCD on the lattice~\cite{Aoki:2013ldr},  and $m_t^{\overline{\rm MS}}(m_t) = 160^{+5}_{-4}$~GeV~\cite{pdg}, 
we get~\footnote{To evaluate $\Delta m_{B_s}^{\rm SM}$ we also used $\eta_B=0.55(1)$~\cite{Buras:1990fn}, and $S_0(x_t)=2.25^{+11}_{-09}$, the Inami-Lim function at $x_t=m_t^2/m_W^2$.}
\bea
\Delta m_{B_s}^{\rm SM}= 17.3\pm 1.7\ {\rm ps}^{-1},
\eea
which is in excellent agreement with the measured $\Delta m_{B_s} =
17.7(2) {\rm ps}^{-1}$~\cite{pdg}. With the values of
$C_{10}^{\prime}$ determined in the previous Section, we see that
Eq.~(\ref{eq:dm}) leads to a very loose upper bound for $m_\Delta$.
For example, for $\re [C_{10 }^{\prime}] \in [0.15, 0.35]$, we get the
upper bound of the order $100$~TeV.

\subsection{Impact of  $(3,2,1/6)$  leptoquark on $B\to K \nu \bar \nu$}
In the presence of leptoquark $\Delta$ the pair of neutrinos in the final
state of $B \to K \nu \bar \nu$ may be in any
flavor combination. In order to encompass such a possibility we must
extend the effective Hamiltonian of Ref.~\cite{Altmannshofer:2009ma}
to account for the disparity in neutrino flavors:
\begin{equation}
{\cal H}_\mrm{eff} = -\frac{ 4 G_F}{ {\sqrt 2}} V_{tb} V_{ts}^* (C_L^{ij} {\cal O}^{ij}_L +C_R^{ij} {\cal O}^{ij}_R) \,.
\end{equation}
The operators are defined as
$ {\cal O}^{ij}_{L,R} =\frac{e^2}{16 \pi^2}(\bar s \gamma_\mu P_{L ,R}
b) (\bar \nu_i \gamma^\mu (1-\gamma_5) \nu_j)$.
The authors of~\cite{Altmannshofer:2009ma} found that in the SM
the Wilson coefficient at next-to-leading order in QCD is
\begin{equation}
C_L^\mrm{SM} \equiv C_L^{ii} = -6.38\pm 0.06\,,\qquad \textrm{(no sum
  over $i$ implied)}.
\end{equation}
If the leptoquark state $(3,2,1/6)$ is present then it will manifest
itself in $B \to K \nu\bar\nu$ through right-handed operators:
\begin{equation}
 C_R^{ij} = -\frac{1}{N}\,\frac{(VY)_{ib} (VY)^*_{js}}{4
   m_\Delta^2}\,,\qquad N \equiv \frac{G_F V_{tb} V_{ts}^* \alpha}{\sqrt{2}\pi}\,.
\end{equation}
Here $V$ denotes the PMNS matrix. The experimentally accessible decay
width of $B \to K \nu \bar\nu$ is a sum of partial widths of $B \to K
\nu^i \bar \nu^j$. The amplitudes are proportional to
the sum of the SM and leptoquark contribution and the two will interfere in
the $B \to K \nu \bar\nu$ decays width as
\begin{equation}
\label{eq:sum}
  \begin{split}
  \Gamma (B \to K \nu \bar \nu) &\sim \sum_{i,j=1}^3
                                   \left|\delta_{ij}  C_L^\mrm{SM} +
                                   C_R^{ij}\right|^2\\
&=3 |C_L^\mrm{SM}|^2 + |C_{10}^{\prime }|^2 - 2 \re[C_L^{\mrm{SM}*}C_{10}^{\prime }]\,.    
  \end{split}
\end{equation}
$C_{10}^{\prime}$ is the Wilson coefficient of $b \to s \mu^+ \mu^-$ that we obtained from the fits to experimental data in the previous Section. 
Last line of Eq.~(\ref{eq:sum}) was obtained by applying the unitarity of matrix
$V$, and assuming that $Y_{\mu b}$
and $Y_{\mu s}$ are the only non-zero elements of the matrix $Y$. Finally, the $q^2$-spectrum of 
this decay reads,
\begin{equation}
\label{eq:Brnu}
  \frac{\!\mrm{d}\Gamma}{\mrm{d}q^2}(B \to K \nu \bar
  \nu)=\frac{|N|^2}{384 \pi^3 m_B^3} \, f_+(q^2) \left[\lambda(m_B^2, m_K^2,
  q^2)\right]^{3/2} \biggl(3 |C_L^\mrm{SM}|^2 + |C_{10}^{\prime }|^2 - 2
  \re[C_L^{\mrm{SM}*} C_{10}^{\prime }]  \biggr)\,,
 \end{equation}
where $q^2$ in this case stands for the invariant mass of
the neutrino pair.
 Notice that the above expression, for $C_{10}^{\prime } = 0$, confirms Eq.~(2.14) of
Ref.~\cite{Altmannshofer:2009ma}. The expression~\eqref{eq:Brnu} can be recast into a product
of the SM $q^2$-spectrum and a correction factor,
\begin{equation}
  \label{eq:nucorrection}
1.01  < \left[1 + \frac{1}{3} \left|C_{10}^{\prime }/C_L^\mrm{SM}\right|^2 - \frac{2}{3}\
 \re[C_{10}^{\prime}/C_L^\mrm{SM}]\right] < 1.05 \ ,
\end{equation}
where its lower and upper bounds have been derived from the $1\sigma$
region of $C_{10}^{\prime }$, obtained in the previous Section. We learn
that the $\mc{B}(B \to K \bar \nu \nu)$ will increase by at most $5\,\%$ if
 leptoquark $\Delta$ is present.

\section{Summary and conclusion\label{CONCL}}

In this paper we discussed a possibility of constraining a scenario of New Physics affecting the $b\to s\mu^+\mu^-$ decays through coupling with the operators ${\cal O}_{9,10}^\prime$. Such a scenario is explicitly verified in a model with a light scalar 
leptoquark state, $\Delta$, carrying the quantum numbers $(3,2,1/6)$ of the Standard Model gauge group. In this scenario, $C_9^\prime = - C_{10}^\prime$ is specific for the muons in the final state. In the leptoquark model discussed in this paper, $C_{10}^\prime$ is related to ${Y_{\mu b} Y_{\mu s}^*}/m_\Delta^2$. From the currently available experimental data on 
$ {\cal B}(B_s\to \mu^+\mu^-)$ and $ {\cal B}(B\to  K \mu^+\mu^-)_{[15,22] \gev^2}$ we were able to constrain $\re [C_{10}^\prime]$ and $\im [C_{10}^\prime]$, which are then used to compute $R_K=  {\cal B}(B\to  K \mu^+\mu^-)/ {\cal B}(B\to  K e^+ e^-)_{[1,6] \gev^2}=0.88(8)$. We find a good  agreement with a recent experimental result, $R_K^{\rm LHCb}=0.75(14)$. 

After having passed this test we focused on $B\to  K^\ast \ell^+ \ell^-$ decays. Within our scenario, and the range of $C_{10}^\prime$ obtained from $ {\cal B}(B_s\to \mu^+\mu^-)$ and $ {\cal B}(B\to  K \mu^+\mu^-)_{[15,22] \gev^2}$, we predict  $R_{K^\ast}=  {\cal B}(B\to  K^\ast \mu^+\mu^- )/ {\cal B}(B\to  K^\ast e^+ e^-)_{[1,6] \gev^2}=1.11(8)$, and $R_{K^\ast}/R_K=1.27(10)$. Similarly, in this scenario the ratio of forward-backward asymmetries becomes different from unity. In particular, we find 
$R_{\rm fb}= ({A_{\rm fb}^\mu /A_{\rm fb}^e })_{ [4-6]\ \gev^2} = 0.8(1)$. 
Furthermore, we checked that a combination of coefficients of the angular distribution of $B\to  K^\ast \mu^+\mu^-$, known as $P_5^\prime$, and weighed over a specific bin of $q^2$'s, indeed becomes smaller 
than its value predicted in the Standard Model. However, it cannot explain a very low value of $| \langle P_5^\prime \rangle |_{[4.3 - 8.68]}$ measured at LHCb, for 
which the Standard Model prediction is still a subject to controversies mainly related to the issue of treatment of the charm quark loops.  

Finally, in the leptoquark model our constraints on the Wilson coefficient $ C_{10}^\prime$ can have impact on other physical processes. We checked, in particular, that 
the contribution to the frequency of oscillation in the $B_s-\overline B_s$ system is insignificant, and that only up to five percent enhancement of $ {\cal B}(B\to K\nu\bar \nu)$ can be obtained. 

\begin{acknowledgments}
S.F. and N.K. acknowledge support of the Slovenian Research Agency.
\end{acknowledgments}

\appendix

\section{$B\to K^{(\ast)}$ form factors\label{sec:app}}

For completeness we remind the reader of the standard parameterization of the $B\to K^{(\ast )} \ell^+\ell^-$ hadronic matrix elements in terms of  the relevant form factors, 
\begin{align}
\langle K(k)| \bar{s} \gamma_{\mu} b |B(p)\rangle &=  \left[ (p +
  k)_\mu - {m_B^2 - m_K^2 \over q^2} q_\mu \right] f_{+} \qq 
+ {m_B^2-m_K^2 \over q^2} q_{\mu} f_{0} \qq \, , \nn \\
\langle K(k)| \bar{s}\sigma_{\mu\nu}b |B(p) \rangle &= -i \left( p_\mu k_\nu  - p_\nu k_\mu \right) \frac{2 f_T\qq}{m_B + m_K} \, ,
\end{align}
\bea\label{eq:ffVA}
\langle \bar K^\ast (k) | \bar s\gamma_\mu(1-\gamma_5) b | \bar B(p)\rangle  &=&  \varepsilon_{\mu\nu\rho\sigma}\varepsilon^{*\nu} p^\rho k^\sigma\,
\frac{2V(q^2)}{m_B+m_{K^\ast}} - i \varepsilon^\ast_\mu (m_B+m_{K^\ast}) A_1(q^2)\nonumber \\
&& \hspace*{-32mm}+ i (p+k)_\mu (\varepsilon^\ast \cdot q)\, \frac{A_2(q^2)}{m_B+m_{K^\ast}}   +  i q_\mu (\varepsilon^\ast \cdot q) \,\frac{2m_{K^\ast}}{q^2}\,
\left[A_3(q^2)-A_0(q^2)\right], 
\eea
\bea\label{eq:ffT}
\langle \bar K^\ast(k) | \bar s \sigma_{\mu\nu} q^\nu (1+\gamma_5) b |
\bar B(p)\rangle &=& 2 i\varepsilon_{\mu\nu\rho\sigma} \varepsilon^{*\nu}
p^\rho k^\sigma \,  T_1(q^2)\nonumber\\
& & + \left[ \varepsilon^\ast_\mu
  (m_B^2-m_{K^\ast}^2) - (\varepsilon^\ast \cdot q) \,(2p-q)_\mu \right] T_2(q^2) \nonumber\\
& &+  
(\varepsilon^\ast \cdot q) \left[ q_\mu - \frac{q^2}{m_B^2-m_{K^\ast}^2}\, (p+k)_\mu
\right]T_3(q^2),
\eea
where $2m_{K^\ast} A_3(q^2) = (m_B+m_{K^\ast})  A_1(q^2) - (m_B-m_{K^\ast})   A_2(q^2)$, and $T_1(0) = T_2(0)$. In the limit of massless lepton, the $q^2$-dependent functions entering eqs.~(\ref{eq:GammaBK},\ref{distr-1}), relevant for the present study, read
\bea
&&a_\mu(q^2)= \frac{3}{2}{\cal F} \,,\qquad  I_2^s = \frac{1}{4}\left( |A_\perp^{L,R}|^2 + |A_\parallel^{L,R}|^2 \right), \nn\\
&& I_2^c = -  |A_0^{L,R}|^2 ,  \qquad I_6^s = 2  \re\left( A_\perp^{L} A_\parallel^{L\ \ast} - A_\perp^{R} A_\parallel^{R\ \ast}  \right) \,,
\eea
where
\bea \label{ampl-F}
{\cal F}(q^2)  &=&  N \lambda(q^2) \left[ 
\biggl|(C_9 + C_9^{\prime}) f_+(q^2) + {2 m_b\over m_B + m_{K} } (C_7 + C_7^{\prime}) f_T(q^2) \biggr|^2
+ \biggl|(C_{10} + C_{10}^{\prime}) f_+(q^2)\biggr|^2
 \right]
,  \nn 
\eea
\bea \label{ampl-A}
A_{\perp}^{L,R}(q^2)  &=&  \sqrt{ N q^2 \lambda(q^2)} \bigg[ \frac{2m_b}{q^2} (C_7 + C_7^{\prime})\ T_1(q^2)  +
\left[ (C_9 + C_9^{\prime}) \mp (C_{10} + C_{10}^{\prime}) \right] { V(q^2) \over m_B + m_{K^\ast} }
\bigg],  \nn \\
A_{\parallel}^{L,R}(q^2)  & =&  - \sqrt{  N q^2 }(m_B^2 - m_{K^\ast}^2) \bigg[ 
\frac{2 m_b}{q^2} (C_7- C_7^{\prime}) \ T_2(q^2)  + \left[ (C_9 - C_9^{\prime}) \mp (C_{10} - C_{10}^{\prime}) \right] 
{A_1(q^2)\over m_B-m_{K^\ast}}
\bigg]\,,  \nn \\
 A_0^{L,R}(q^2) &= & -\frac{\sqrt{ N}}{2m_{K^\ast}  } \biggl\{  \left[ \left( C_9 - C^{\prime}_9 \right)   \mp   \left(C_{10}- C^{\prime }_{10} \right) \right]\times \biggr.\nn\\
		    & & \qquad\left[ \left( m_B^2 -m_{K^\ast}^2 - q^2\left) \right( m_B +m_{K^\ast} \right) A_1(q^2) - \lambda(q^2)  \frac{A_2(q^2)}{m_B + m_{K^\ast}} \right] \\
		     &&\qquad \left.+ 2 m_b \left(C_7- C^{\prime}_7 \right) \left[ \left( m_B^2 + 3 m_{K^\ast}^2 - q^2\right) T_2(q_2) - \frac{\lambda(q^2)}{m_B^2 - m_{K^\ast}^2} T_3(q^2) \right] \right\}\,,\nn
 \eea
and $N= |V_{tb}V_{ts}^\ast|^2 \ \lambda^{1/2}(q^2) \  G_F^2 \alpha^2/( 3072 \pi^5 m_B^3)$,  $ \lambda(q^2) = [q^2 - (m_B+m_{K^\ast})^2]\ [q^2 - (m_B-m_{K^\ast})^2]$. 
In the same massless lepton limit, $a_\mu=-c_\mu$, $I_1^s=3I_2^s$ and $I_1^c=-I_2^c$, so that $d\Gamma(B\to K \mu\mu)/dq^2 =4 a_\mu/3$,  and 
$d\Gamma(B\to K^\ast \mu\mu)/dq^2 =4I_2^s-I_2^c = |A_\perp^{L,R}|^2 + |A_\parallel^{L,R}|^2 + |A_0^{L,R}|^2$.  The two transverse asymmetries discussed in the text are defined as $A_T^{(2)}(q^2)= I_3/(2 I_2^s)= ( |A_\perp |^2 - |A_\parallel |^2)/( |A_\perp |^2 + |A_\parallel |^2)$, and $A_T^{(\mathrm{Im})}(q^2) =  I_9/(2 I_2^s)=- 2 \im( A_\parallel A_\perp^\ast )/( |A_\perp |^2 + |A_\parallel |^2)$. 
In the computation of decay amplitudes we rely on the full QCD form factors computed on the lattice (in case of $B\to K\ell\ell$) or by means of the QCD sum rules near the light cone (in case of $B\to K^\ast\ell\ell$) and to the 
standard (universal) Operator Product Expansion. We did not rely on the effective theory approaches such as QCD factorization because we do not know how to reliably compute the relevant form factors and include all power corrections. At the level of present accuracy however, both approaches lead to compatible results.

\bibliography{refs}

\end{document}